\documentclass[prd,aps,nofootinbib]{revtex4}
\usepackage{amsmath,graphicx,color}

\begin{document}
 \title{The CKM suppressed $B(B_s) \to \bar D_{(s)}P,\bar D_{(s)}V,\bar D_{(s)}^*P,\bar D_{(s)}^*V$
  decays in perturbative QCD approach}

 \author{Hao Zou$^{a}$, Run-Hui Li$^{a,b}$, Xiao-Xia Wang$^a$, and Cai-Dian L\"{u}$^{a,c}$}

 \affiliation{
 \it $^a$ Institute of High Energy Physics, P.O. Box 918(4), Beijing 100049, People's Republic of China\\
 \it $^b$  School of Physics, Shandong University, Jinan 250100, People's Republic of China \\
 \it $^c$ Theoretical Physics Center for Science Facilities, Beijing 100049, People's Republic of China}

 \date{\today}
 \begin{abstract}
Although the two-body charmed decays $B_{(s)} \to  \bar
D_{(s)}^{(*)}P$ and $\bar D_{(s)}^{(*)}V$, where $P$($V$) denotes a
light pseudoscalar(vector) meson, are CKM suppressed comparing with
the $B_{(s)} \to    D_{(s)}^{(*)}P$ and $  D_{(s)}^{(*)}V$ decays,
they are important in the CKM angle $\gamma=\phi_3$ extraction
method. We investigated   these decays in the perturbative QCD
approach to the leading order of  $m_D/m_B$  and
$\Lambda_{\rm{QCD}}/m_D$ expansion. We  find that the
nonfactorizable emission diagrams and the annihilation diagrams are
not negligible in many of these channels. The numerical results show
that most channels have   branching ratios with an order of
$10^{-6}$ or $10^{-7}$. The ratio needed for the CKM angle $\gamma$
extraction is estimated as $r = \frac{|A(B^- \to \bar{D}^0
K^-)|}{|A(B^- \to D^0 K^-)|} = 0.092_{-0.003-0.003}^{+0.012+0.003},$
which is too small for the experiments. Some   of the $ B_{(s)}\to
\bar D_{(s)}^*V$ decays have a very large transversely polarized
contribution that can reach 80\%.

 \end{abstract}
 \maketitle

 \section{introduction}
The study of B physics   plays an important role in precise test of
the standard model,  extraction of the
Cabbibo-Kobayashi-Maskawa(CKM) matrix elements, searching for the
origin of CP violation and new physics signals and even discovery of
new hadronic states. The hadronic B  decays offer  an opportunity to
understand the nonperturbative QCD. After years of hard work,
theorists develop many approaches to deal with the nonleptonic
decays of B mesons, such as QCD factorization approach \cite{QCDF},
soft collinear effective theory \cite{SCET}, perturbative QCD
approach(PQCD)\cite{pQCD}, QCD sum rules\cite{QCDSR}, light cone sum
rules\cite{LCSR}. At the experimental side, the two B factories have
accumulated a great amount of data, which can be used to either test
various theoretical approaches or carry on new physics
investigations.

The two body charmed decays of B mesons $B_{(s)} \to D_{(s)}^{(*)}P$
and $  D_{(s)}^{(*)}V$, which are important in the extraction of the
CKM angles \cite{CKMextraction}, have been investigated in the PQCD
approach \cite{BtoDprevious1, BtoDprevious,BtoD1}. These channels
are induced by the $b\to c$ transitions, which are CKM favored.
However, the method of CKM angle $\gamma$ extraction
\cite{Gronau:1998vg} also requires another category of charmed meson
B decays, which are induced by $b\to u$ transition. The interference
between the $b\to c$ and $b\to
 u$ transitions gives the measurement of the CKM angle $\gamma$.
 These $b\to u$ decays are CKM $|V_{ub}/V_{cb}|$ suppressed, thus will have
smaller branching ratios. In this paper, we will investigate these
$B(B_s) \to \bar D_{(s)}P,\bar D_{(s)}V,\bar D_{(s)}^*P,\bar
D_{(s)}^*V$ decays in the PQCD approach.

Unlike the collinear factorization in the QCD factorization approach
and soft-collinear effective theory, the $k_T$ factorization is
utilized in the PQCD approach. In this approach, the transverse
momentum of valence quarks in the mesons are kept to avoid the
endpoint singularity.  Therefore, only in this factorization method,
one can calculate the color suppressed channels as well as the color
allowed channels in charmed B decays. The conventional
non-calculable annihilation type decays are also calculable in the
PQCD approach, which is proved to be the dominant strong phase in
$B$ decays for the direct CP asymmetry \cite{dir}.  In the PQCD
approach, the most important uncertainties come from the hadronic
wave functions. We will use the same hadronic parameters determined
from the charmed B decays induced by $b\to c$ transitions
\cite{BtoD1} to reduce uncertainties. The numerical results show
that the $b\to u$ transition decays are indeed heavily suppressed
comparing with those $b\to c$ transitions. Thus the CKM angle
measurement method do face difficulty experimentally.

The paper is organized as follows: In Sec.~\ref{sec:formulae} we
list the analytic formulae needed in the calculation, including the
Hamiltonian and definition of the momenta, PQCD factorization
formulae of all the diagrams, and the expressions of the amplitudes
for all the decay channels. Sec.~\ref{sec:numerical} contains the
numerical results we obtain, and some discussions. A brief summary
is given in Sec.~\ref{sec:summary}. The wave functions, decay
constants and some functions that appear in PQCD approach are put in
the appendices.

\section{analytic formulae for decay amplitudes}
\label{sec:formulae}

In B meson weak decays, there are three natural energy scales
involved: W boson mass, b quark mass scale and the hadronic scale
$\Lambda_{QCD}$. The electroweak physics higher than W boson mass
can be perturbatively calculated. The physics between b quark mass
and W boson mass can also be calculated using the renormalization
group equation. Both of these two contributions are included in the
well calculated effective Hamiltonian of the four quark operators.
The physics below b mass scale is complicated, where we have to
utilize the factorization theorem to factorize the non-perturbative
contribution out. In this purpose, we do the $1/m_b$ expansion in
the soft collinear effective theory \cite{SCET}. Unfortunately,
there are some contributions, which produce endpoint singularity. In
order to deal with this singularity, usually there are two ways of
doing it. One of them is  the PQCD approach, in which we keep the
transverse momentum of the valence quark. By doing this $k_T$
factorization, a new series of double logs are generated. Using the
renormalization group equation, we resum these logs to give a
Sudakov factor, which suppresses the endpoint contribution. Finally
the decay amplitude becomes
\begin{eqnarray}
{\cal M}= \int d^4{{k}}_{1}d^4{ {k}}_{2}d^4{ {k}}_{3}\phi_B({
{k}}_{1},t) T_H({ {k}}_{1},{ {k}}_{2},{ {k}}_{3},t) \phi_{P_2}({
{k}}_{2},t)\phi_{P_3}( { {k}}_{3},t) e^{S({k_i,t})},
\end{eqnarray}
where $T_H$ is the hard part that is perturbatively calculable, and
$\phi_M$ are the hadronic meson wave functions that is
non-perturbative. The Sudakov factor $e^{S({k_i,t})}$ resulting from
the resummation of double logarithm, relate the perturbative and
non-perturbative dynamics.

In charmed $B$ decays, we have one more intermediate energy scale,
the D meson mass. Therefore, one encounters another expansion series
of $m_D/m_B$. The factorization is only approved at the leading of
$m_D/m_B$ expansion \cite{BtoDprevious1,BtoDprevious}, so that we
will keep only the leading order contribution numerically, unless
explicitly mentioned.

 \subsection{Effective Hamiltonian and Kinematics}

For the processes considered, only tree operators contribute, and
the effective Hamiltonian is given by
 \begin{equation}
 \label{effective_hamiltanian}
 {\cal
 H}_{eff}=\frac{G_F}{\sqrt{2}}V_{ub}V^*_{cD}\left[C_1(\mu)O_1(\mu)+C_2(\mu)O_2(\mu)\right],
 \end{equation}
 with $D=d,\;s$, and
 \begin{eqnarray}
 O_1=(\bar u_{\alpha}b_{\beta})_{V-A}(\bar D_{\beta}c_{\alpha})_{V-A}\;\;,
 O_2=(\bar u_{\alpha}b_{\alpha})_{V-A}(\bar
 D_{\beta}c_{\beta})_{V-A}\;\;.
 \end{eqnarray}
 Here, $\alpha$ $\beta$ are the color indices,
 $(\bar q_1q_2)_{V-A}\equiv\bar
 q_1\gamma^{\mu}(1-\gamma^5)q_2$. And the $V_{ub}V^*_{cD}$ are the corresponding CKM matrix elements.

Sandwiching the above Hamiltonian between the initial and final
state mesons£¬ and factorizing the matrix elements, the combinations
of the Wilson coefficients usually  appear. Conventionally, they are
defined as
 \begin{eqnarray}
 a_1=C_2+C_1/3\;,\;a_2=C_1+C_2/3,\label{eq:wilsona}
 \end{eqnarray}
where,  $a_1$ and $a_2$ correspond to the color favored and color
suppressed contribution,  respectively. The light-cone coordinates
are used in this paper, with which the vector $V$ is expressed as
$V=(\frac{V^0+V^3}{\sqrt{2}},\frac{V^0-V^3}{\sqrt{2}},V_{\perp})$,
where $V_{\perp}=(V^1,V^2)$. The momenta of $B$, $D$ and the light
mesons are respectively $P_1$, $P_2$ and $P_3$, which are defined as
 \begin{eqnarray}
 P_1=\frac{m_B}{\sqrt{2}}(1,1,\textbf{0}_{\perp})\;,\;
 P_2=\frac{m_B}{\sqrt{2}}(1,r^2,\textbf{0}_{\perp})\;,\;
 P_3=\frac{m_B}{\sqrt{2}}(0,1-r^2,\textbf{0}_{\perp})\;,
 \end{eqnarray}
with $r=m_D/m_B$. The momenta of the light quarks in B and D mesons
are denoted by $k_1$ and $k_2$, respectively, whereas $k_3$
represents the momentum of the quark in the light meson. Both
$k_1^+$ and $k_1^-$ contribute, but considering $k_2 \sim
\mbox{O}(\Lambda)$ and $k_1 \sim \mbox{O}(x_1 m_B)$, we drop the
term $k_1 \cdot k_1$ for the emission diagrams and $k_1 \cdot k_2$
for the annihilation diagrams. The effect is equal to dropping
$k_1^-$ for emission diagrams and dropping $k_1^+$ for annihilation
diagrams. Their explicit expressions are
 \begin{eqnarray}
k_1&=&(x_1\frac{m_B}{\sqrt{2}},0,\textbf{k}_{1\perp})\;\mbox{for
emission diagrams,}\nonumber\\
k_1&=&(0,x_1\frac{m_B}{\sqrt{2}},\textbf{k}_{1\perp})\;\mbox{for annihilation diagrams\;,}\nonumber\\
k_2&=&(x_2\frac{m_B}{\sqrt{2}},0,\textbf{k}_{2\perp})\;,\;
k_3=(0,x_3\frac{(1-r^2)m_B}{\sqrt{2}},\textbf{k}_{3\perp})\;.
 \end{eqnarray}
 Here, $x_1$, $x_2$ and $x_3$ are the momentum fractions, and
 $\textbf{k}_{1\perp}$, $\textbf{k}_{2\perp}$ and
 $\textbf{k}_{3\perp}$ are the transverse momenta of the quarks.

  \begin{figure}
 \begin{center}
 \includegraphics[scale=0.8]{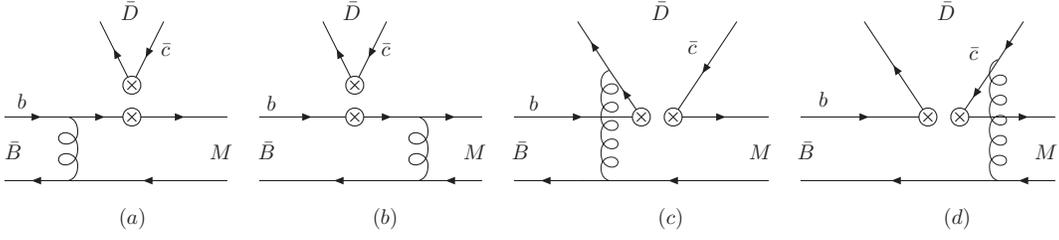}
 \caption{Emission diagrams in pQCD approach for $B\to\bar DP$ decays.}
 \label{fig:int}
 \end{center}
 \end{figure}

  \begin{figure}
 \begin{center}
 \includegraphics[scale=0.8]{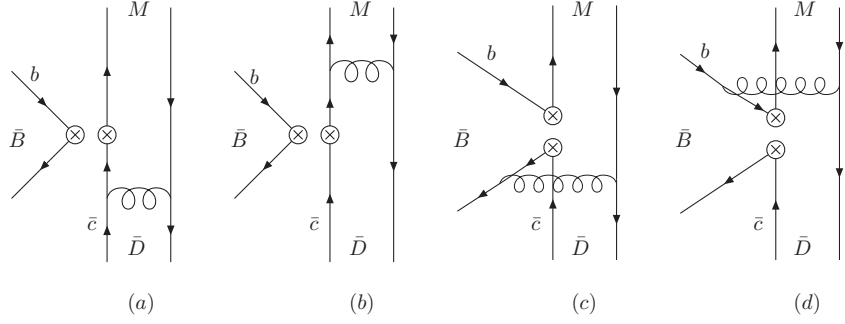}
 \caption{Annihilation diagrams in pQCD approach for $B\to\bar DP$ decays.}
 \label{fig:exc}
 \end{center}
 \end{figure}

\subsection{Factorization formulae of $B \to \bar DP$}

The contributions to the $B \to \bar D P$ processes can be divided
into two types: The emission diagrams(see Fig. \ref{fig:int},
denoted by a subscript $int$ in the following formulae), in which
the light quark in the B mesons is hadronized into one of the final
state mesons as a spectator, and the annihilation diagrams(see Fig.
\ref{fig:exc}, denoted by a subscript $exc$ in the following
formulae) with no spectator quarks. The first two diagrams of Fig.
\ref{fig:int} are the factorizable diagrams, which can be factorized
as a product of the decay constant of $\bar D$ meson and a B to
light meson transition form factor. Summing the expressions of these
two diagrams together, we obtain
  \begin{eqnarray}
 \xi_{\rm int}(a_i)&=&8\pi C_Ff_D \int_0^1
 dx_1dx_3\int_0^{1/\Lambda}b_1db_1b_3db_3 \phi_B(x_1,b_1)
 \nonumber \\
 & &\times \{[(2-x_3) \phi_P(x_3) - r_0 (1- 2x_3)
 (\phi_P^p(x_3)- \phi_P^T(x_3))]
 \nonumber\\
 & &\times a_i(t_i^{(1)})E_i(t_i^{(1)})h_i(x_1,(1-x_3)(1-r^2),b_1,b_3)S_t(x_3)
 \nonumber \\
 & &\ \ + 2r_0 \phi_P^p(x_3)a_i(t_i^{(2)}) E_i(t_i^{(2)})
 h_i(1-x_3,x_1(1-r^2),b_3,b_1)S_t(x_1)\}\;,
 \label{int}
 \end{eqnarray}
where $a_i$ is the corresponding combination of Wilson coefficients
and $r_0=m_0/m_B$, with $m_0$ as the chiral mass of the pseudoscalar
mesons.  The expressions of the PQCD factorization functions $h_j$,
jet function $S_t(x)$ and $E_j(t_j^l)$ and scales $t_j^l$, with
$j=i$, $a$, $d$, $f$ and $l=1$ or $2$, are listed in Appendix
\ref{appendix:pQCDfunctions}.

In the factorized diagrams of annihilation contributions(the first
two diagrams of Fig. \ref{fig:exc}), the B meson is factorized out.
And the combination of these two diagrams give
 \begin{eqnarray}
 \xi_{\rm exc}(a_i)&=&8\pi C_Ff_B \int_0^1
 dx_2dx_3\int_0^{1/\Lambda}b_2db_2b_3db_3 \phi_{D}(x_2,b_2)
 \nonumber \\
 & &\times \left[(-x_2\phi_P(x_3)-2r_0r(1+x_2)\phi_P^p(x_3))a_i(t_a^{(1)})
 E_a(t_a^{(1)}) h_a(1-x_3,x_2(1-r^2),b_3,b_2)S_t(x_3)\right.
 \nonumber \\
 && \left.+((1-x_3)\phi_P(x_3)+r_0r((2x_3-1)\phi_P^T(x_3)+(3-2x_3)\phi_P^p(x_3)))a_i(t_a^{(2)})E_a(t_a^{(2)})\right.\nonumber\\
 &&\left.\times h_a(x_2,(1-x_3)(1-r^2),b_2,b_3)S_t(x_2)\right]\;.
 \label{exc}
 \end{eqnarray}
The last two diagrams of Fig. \ref{fig:int} and \ref{fig:exc} are
nonfactorizable diagrams. Generally, if the two final state mesons
are both light ones, the nonfactorizable contributions of the
emission diagrams are very small, because of the cancelation between
the two nonfactorizable diagrams. While for $\bar D$ mesons, since
the heavy $\bar c$ quark and the light quark is not symmetric, the
nonfactorizable emission diagrams give remarkable contributions. The
expression of the nonfactorizable emission contributions is
  \begin{eqnarray}
 {\cal M}_{\rm int}(a_i)&=& 16\pi\sqrt{2N_c} C_F \int_0^1
 [dx]\int_0^{1/\Lambda}b_1 db_1 b_2 db_2 \phi_B(x_1,b_1)\phi_D(x_2,b_2)
 \nonumber \\
 & &\times
 \left[\left(x_2\phi_P(x_3)+r_0(x_3-1)(\phi_P^p(x_3)+\phi_P^T(x_3))\right)a_i(t_d^{(1)})
 E_d(t_d^{(1)})h^{(1)}_d(x_i,b_i)\right.\nonumber\\
 &
 &\left.+\left((x_3+x_2-2)\phi_P(x_3)+r_0(1-x_3)(\phi_p^p(x_3)-\phi_P^T(x_3))\right)a_i(t_d^{(2)})
 E_d(t_d^{(2)})h^{(2)}_d(x_i,b_i)\right]\;,
 \label{md}
 \end{eqnarray}
 and that of the nonfactorizable annihilation contributions is
 \begin{eqnarray}
 {\cal M}_{\rm exc}(a_i)&=& 16 \pi\sqrt{2N_c} C_F \int_0^1
 [dx]\int_0^{1/\Lambda}b_1 db_1 b_2 db_2
 \phi_B(x_1,b_1)\phi_{D}(x_2,b_2)
 \nonumber \\
 & &\times [((x_3-1)\phi_P(x_3)-r_0r((x_2-x_3+3)\phi_P^p(x_3)+(x_2+x_3-1)\phi_P^T(x_3)))a_i(t_f^{(1)})
 E_f(t_f^{(1)})h^{(1)}_f(x_i,b_i)\nonumber\\
 & & +(x_2\phi_P(x_3)-r_0r((x_3-x_2-1)\phi_P^p(x_3)+(x_2+x_3-1)\phi_P^T(x_3)))a_i(t_f^{(2)}) E_f(t_f^{(2)})h^{(2)}_f(x_i,b_i) ]\;.
 \label{mf}
 \end{eqnarray}
\subsection{Amplitudes for $ B \to \bar D P$}
With the functions obtained in the above, the amplitudes of $22$ $
B \to \bar D P$ decay channels can be given by
\begin{eqnarray}
 {\cal A}(B^-\to \bar D^0 \pi^-) &=& \frac{G_F} {\sqrt{2}}
 m_B^4 V_{ub}V_{cd}^* \left(\xi_{int}(a_2)+{\cal M}_{int}(C_2/3)+\xi_{exc}(a_1)+{\cal M}_{exc}(C_1/3)\right) ,\label{amp1}\\
 {\cal A}(B^-\to \bar D^0 K^-) &=& \frac{G_F} {\sqrt{2}}
 m_B^4 V_{ub}V_{cs}^* \left(\xi_{int}(a_2)+{\cal M}_{int}(C_2/3)+\xi_{exc}(a_1)+{\cal M}_{exc}(C_1/3)\right) ,\\
 {\cal A}(B^-\to  D^- \pi^0) &=& \frac{G_F} {\sqrt{2}}
 m_B^4 V_{ub}V_{cd}^*\frac{1}{\sqrt{2}}\left(\xi_{int}(a_1)+{\cal M}_{int}(C_1/3)-\xi_{exc}(a_1)-{\cal M}_{exc}(C_1/3)\right) ,\\
 {\cal A}(B^-\to D^- \bar K^0) &=& \frac{G_F} {\sqrt{2}}
 m_B^4 V_{ub}V_{cs}^* \left(\xi_{exc}(a_1)+{\cal M}_{exc}(C_1/3)\right) ,\\
 {\cal A}(B^-\to D^- \eta_{n\bar n}) &=& \frac{G_F} {\sqrt{2}}
 m_B^4 V_{ub}V_{cd}^* \frac{1}{\sqrt{2}}\left(\xi_{int}(a_1)+{\cal M}_{int}(C_1/3)+\xi_{exc}(a_1)+{\cal M}_{exc}(C_1/3)\right) ,\\
 {\cal A}(B^-\to D_s^- \pi^0) &=& \frac{G_F} {\sqrt{2}}
 m_B^4 V_{ub}V_{cs}^* \frac{1}{\sqrt{2}}\left(\xi_{int}(a_1)+{\cal M}_{int}(C_1/3)\right),\\
 {\cal A}(B^-\to D_s^- K^0) &=& \frac{G_F} {\sqrt{2}}
 m_B^4 V_{ub}V_{cd}^* \left(\xi_{exc}(a_1)+{\cal M}_{exc}(C_1/3)\right) ,\\
 {\cal A}(B^-\to D_s^- \eta_{n\bar n}) &=& \frac{G_F} {\sqrt{2}}
 m_B^4  V_{ub}V_{cs}^* \frac{1}{\sqrt{2}}\left(\xi_{int}(a_1)+{\cal M}_{int}(C_1/3)\right),\\
 {\cal A}(B^-\to D_s^- \eta_{s\bar s}) &=& \frac{G_F} {\sqrt{2}}
 m_B^4 V_{ub}V_{cs}^* \left(\xi_{exc}(a_1)+{\cal
 M}_{exc}(C_1/3)\right),\\
 {\cal A}(\bar B^0 \to \bar D^0 \pi^0) &=& \frac{G_F} {\sqrt{2}}
 m_B^4 V_{ub}V_{cd}^* \frac{1}{\sqrt{2}} \left(-\xi_{int}(a_2)-{\cal M}_{int}(C_2/3)+\xi_{exc}(a_2)+{\cal M}_{exc}(C_2/3)\right), \\
 {\cal A}(\bar B^0 \to \bar D^0 \bar K^0) &=& \frac{G_F} {\sqrt{2}}
 m_B^4 V_{ub}V_{cs}^* \left(\xi_{int}(a_2)+{\cal M}_{int}(C_2/3)\right) ,\\
 {\cal A}(\bar B^0 \to \bar D^0 \eta_{n\bar n}) &=& \frac{G_F} {\sqrt{2}}
 m_B^4 V_{ub}V_{cd}^* \frac{1}{\sqrt{2}} \left(\xi_{int}(a_2)+{\cal M}_{int}(C_2/3)+\xi_{exc}(a_2)+{\cal M}_{exc}(C_2/3)\right) ,\\
 {\cal A}(\bar B^0 \to D^- \pi^+) &=& \frac{G_F} {\sqrt{2}}
 m_B^4 V_{ub}V_{cd}^* \left(\xi_{int}(a_1)+{\cal M}_{int}(C_1/3)+\xi_{exc}(a_2)+{\cal M}_{exc}(C_2/3)\right) ,\\
 {\cal A}(\bar B^0 \to D_s^- \pi^+) &=& \frac{G_F} {\sqrt{2}}
 m_B^4 V_{ub}V_{cs}^*  \left(\xi_{int}(a_1)+{\cal M}_{int}(C_1/3)\right) ,\\
 {\cal A}(\bar B^0 \to D_s^- K^+) &=& \frac{G_F} {\sqrt{2}}
 m_B^4 V_{ub}V_{cd}^* \left(\xi_{exc}(a_2)+{\cal
 M}_{exc}(C_2/3)\right),\\
 {\cal A}(\bar B_s^0 \to \bar D^0 \pi^0) &=& \frac{G_F} {\sqrt{2}}
 m_B^4 V_{ub}V_{cs}^* \frac{1}{\sqrt{2}} \left(\xi_{exc}(a_2)+{\cal M}_{exc}(C_2/3)\right) ,\\
 {\cal A}(\bar B_s^0 \to \bar D^0 K^0) &=& \frac{G_F} {\sqrt{2}}
 m_B^4 V_{ub}V_{cd}^*  \left(\xi_{int}(a_2)+{\cal M}_{int}(C_2/3)\right) ,\\
 {\cal A}(\bar B_s^0 \to \bar D^0 \eta_{n\bar
n}) &=& \frac{G_F} {\sqrt{2}}
 m_B^4 V_{ub}V_{cs}^*\frac{1}{\sqrt{2}} \left(\xi_{exc}(a_2)+{\cal M}_{exc}(C_2/3)\right) ,\\
 {\cal A}(\bar B_s^0 \to \bar D^0 \eta_{s\bar s}) &=& \frac{G_F} {\sqrt{2}}
 m_B^4 V_{ub}V_{cs}^* \left(\xi_{int}(a_2)+{\cal M}_{int}(C_2/3)\right) ,\\
 {\cal A}(\bar B_s^0\to D^- \pi^+) &=& \frac{G_F} {\sqrt{2}}
 m_B^4 V_{ub}V_{cs}^*  \left(\xi_{exc}(a_2)+{\cal M}_{exc}(C_2/3)\right) ,\\
 {\cal A}(\bar B_s^0\to D^- K^+) &=& \frac{G_F} {\sqrt{2}}
 m_B^4 V_{ub}V_{cd}^*  \left(\xi_{int}(a_1)+{\cal M}_{int}(C_1/3)\right) ,\\
 {\cal A}(\bar B_s^0 \to D_s^- K^+) &=& \frac{G_F} {\sqrt{2}}
 m_B^4 V_{ub}V_{cs}^* \left(\xi_{int}(a_1)+{\cal
 M}_{int}(C_1/3) + \xi_{exc}(a_2)+{\cal
 M}_{exc}(C_2/3)\right),\label{amp22}
\end{eqnarray}
where $\eta_{n\bar n}=\frac{1}{\sqrt{2}}(u\bar u+d\bar d)$ and
$\eta_{s\bar s}=s\bar s$. We treat $\eta$ and $\eta^{\prime}$ as the
mixtures of $\eta_{n\bar n}$ and $\eta_{s\bar s}$ with
\begin{equation}
   \left( \begin{array}{c}
    \eta \\ \eta^\prime
   \end{array} \right)
   =\left( \begin{array}{cc}
    \cos\phi & ~-\sin\phi \\
    \sin\phi & \phantom{~-}\cos\phi
   \end{array} \right)
   \left( \begin{array}{c}
    \eta_{n\bar n} \\ \eta_{s\bar s}
   \end{array} \right),
\end{equation}
where $\phi=39.3^\circ\pm1.0^\circ$\cite{etamixing}.

 \subsection{Decay Amplitudes for   $B\to \bar D^*P$ and $B\to \bar D^{(*)}V$  modes}
 \label{sec:amplitudes}

Due to the conservation of the angular momentum, only the
longitudinally polarized vector mesons are generated in the $B\to
\bar DV$ and $B\to \bar D^*P$ decay modes. According to the
similarity between the Lorentz structures of the wave functions for
the pseudoscalar mesons and vector mesons, the emission diagrams'
factorization formulae of $B\to\bar DV$ modes can be obtained by
making the following substitutions in Eq. (\ref{int}) and (\ref{md})
up to the leading order and leading power of $1/m_B$:
 \begin{equation}
 \phi_P \to \phi_V, \; \phi_P^p \to -\phi_V^s, \; \phi_P^T \to
 -\phi_V^t,\; r_0 \to r_V,\; f_P \to f_V.
 \end{equation}
The factorization formula for the annihilation diagrams of
$B\to\bar D V$ decays are listed below:
 \begin{eqnarray}
 \xi_{\rm exc}(a_i)&=&8\pi C_Ff_B \int_0^1
 dx_2dx_3\int_0^{1/\Lambda}b_2db_2b_3db_3 \phi_{D}(x_2,b_2)
 \nonumber \\
 & &\times \left[(-x_2\phi_V(x_3)+2r_Vr(1+x_2)\phi_V^s(x_3))a_i(t_a^{(1)})
 E_a(t_a^{(1)}) h_a(1-x_3,x_2(1-r^2),b_3,b_2)S_t(x_3)\right.
 \nonumber \\
 &&\left.+((1-x_3)\phi_V(x_3)+r_Vr((1-2x_3)\phi_V^t(x_3)-(3-2x_3)\phi_V^s(x_3)))a_i(t_a^{(2)})E_a(t_a^{(2)})\right.\nonumber\\
 &&\left.\times h_a(x_2,(1-x_3)(1-r^2),b_2,b_3)S_t(x_2)\right]\;.\\
{\cal M}_{\rm exc}(a_i)&=& 16 \pi\sqrt{2N_c} C_F \int_0^1
 [dx]\int_0^{1/\Lambda}b_1 db_1 b_2 db_2
 \phi_B(x_1,b_1)\phi_{D}(x_2,b_2)
 \nonumber \\
 & &\times [((x_3-1)\phi_V(x_3)+r_Vr((x_2+x_3-1)\phi_V^t(x_3)+(x_2-x_3+3)\phi_V^s(x_3)))a_i(t_f^{(1)})
 E_f(t_f^{(1)})h^{(1)}_f(x_i,b_i)\nonumber\\
 & & +(x_2\phi_V(x_3)-r_Vr((x_2-x_3+1)\phi_V^s(x_3)-(x_2+x_3-1)\phi_V^t(x_3)))a_i(t_f^{(2)}) E_f(t_f^{(2)})h^{(2)}_f(x_i,b_i) ]\;.
 \end{eqnarray}
 By changing the pseudoscalar nonet to corresponding vector
nonet in Eq.(\ref{amp1})-(\ref{amp22}), one will obtain the decay
amplitudes formula for the corresponding $B\to\bar DV$   mode.

While in order to obtain the factorization formulae for $B\to \bar
D^*P$ mode, the substitutions
 \begin{equation}
 \phi_D \to \phi_D^L,\; f_D \to f_{D^*},\; m_D \to m_{D^*}
 \end{equation}
should be made in Eq.(\ref{int}) and (\ref{md}) for emission
diagrams. The annihilation diagrams' formulae are listed below:
 \begin{eqnarray}
 \xi_{\rm exc}(a_i)&=&8\pi C_Ff_B \int_0^1
 dx_2dx_3\int_0^{1/\Lambda}b_2db_2b_3db_3 \phi_{D}(x_2,b_2)
 \nonumber \\
 & &\times \left[(-x_2\phi_P(x_3)+2r_0r(1-x_2)\phi_P^p(x_3))a_i(t_a^{(1)})
 E_a(t_a^{(1)}) h_a(1-x_3,x_2(1-r^2),b_3,b_2)S_t(x_3)\right.
 \nonumber \\
 & &  \left.+((1-x_3)\phi_P(x_3)+r_0r(\phi_P^T(x_3)+\phi_P^p(x_3)))a_i(t_a^{(2)})E_a(t_a^{(2)})
 h_a(x_2,(1-x_3)(1-r^2),b_2,b_3)S_t(x_2)\right]\;.\\
{\cal M}_{\rm exc}(a_i)&=& 16 \pi\sqrt{2N_c} C_F \int_0^1
 [dx]\int_0^{1/\Lambda}b_1 db_1 b_2 db_2
 \phi_B(x_1,b_1)\phi_{D}(x_2,b_2)
 \nonumber \\
 & &\times [((x_3-1)\phi_P(x_3)-r_0r((1-x_2-x_3)\phi_P^p(x_3)+(1-x_2+x_3)\phi_P^T(x_3)))a_i(t_f^{(1)})
 E_f(t_f^{(1)})h^{(1)}_f(x_i,b_i)\nonumber\\
 & & +(x_2\phi_P(x_3)+r_0r((x_3-x_2-1)\phi_P^T(x_3)+(x_2+x_3-1)\phi_P^p(x_3)))a_i(t_f^{(2)}) E_f(t_f^{(2)})h^{(2)}_f(x_i,b_i) ]\;.
 \end{eqnarray}
 The decay
amplitudes for these modes can be got through changing the $\bar D$
mesons to corresponding $\bar D^*$ mesons in
(\ref{amp1})-(\ref{amp22}).

The situation for $B\to\bar D^*V$ mode is a little more complicated.
Both the longitudinal polarization and the transverse polarization
contribute. The longitudinally polarized factorization formulae are
obtained by making the following substitutions in Eq. (\ref{int})
and (\ref{md}):
 \begin{equation}
 \phi_P \to -\phi_V, \; \phi_P^p \to \phi_V^s, \; \phi_P^T \to
 \phi_V^t,\; r_0 \to r_V,\; f_P \to f_V,\;\phi_D \to \phi_D^L, \; f_D \to f_{D^*},\; m_D \to m_{D^*}.
 \end{equation}
 The annihilation diagrams' formula are
 \begin{eqnarray}
 \xi_{\rm exc}(a_i)&=&-8\pi C_Ff_B \int_0^1
 dx_2dx_3\int_0^{1/\Lambda}b_2db_2b_3db_3 \phi_{D}(x_2,b_2)
 \nonumber \\
 & &\times \left[(-x_2\phi_V(x_3)+2r_Vr(x_2-1)\phi_V^s(x_3))a_i(t_a^{(1)})
 E_a(t_a^{(1)}) h_a(1-x_3,x_2(1-r^2),b_3,b_2)S_t(x_3)\right.
 \nonumber \\
 & &  \left.+((1-x_3)\phi_V(x_3)-r_Vr(\phi_V^t(x_3)+\phi_V^s(x_3)))a_i(t_a^{(2)})E_a(t_a^{(2)})
 h_a(x_2,(1-x_3)(1-r^2),b_2,b_3)S_t(x_2)\right]\;.\\
{\cal M}_{\rm exc}(a_i)&=& -16 \pi\sqrt{2N_c} C_F \int_0^1
 [dx]\int_0^{1/\Lambda}b_1 db_1 b_2 db_2
 \phi_B(x_1,b_1)\phi_{D}(x_2,b_2)
 \nonumber \\
 & &\times [((x_3-1)\phi_V(x_3)-r_Vr((x_2-x_3-1)\phi_V^t(x_3)+(x_2+x_3-1)\phi_V^s(x_3)))a_i(t_f^{(1)})
 E_f(t_f^{(1)})h^{(1)}_f(x_i,b_i)\nonumber\\
 & & +(x_2\phi_V(x_3)+r_Vr((1-x_2-x_3)\phi_V^s(x_3)+(x_2-x_3+1)\phi_V^t(x_3)))a_i(t_f^{(2)}) E_f(t_f^{(2)})h^{(2)}_f(x_i,b_i) ]\;.
 \end{eqnarray}
The longitudinally polarized decay amplitudes ${\cal A}^N$s can be
obtained through replacing the pseudo-scalar mesons by  the
corresponding  light vector mesons and the $\bar D^*$ meson in
(\ref{amp1})-(\ref{amp22}).

The transversely polarized amplitude of $B\to \bar D^*V$ mode can be
decomposed as
 \begin{eqnarray}
 {\cal
 A}^T(\epsilon^{T*}_D,\epsilon^{T*}_V)=\frac{G_F}{\sqrt{2}}m_B^4V_{ub}V^*_{cD}
 \bigg[i(\epsilon^{T*}_D\cdot\epsilon^{T*}_V)A^s + \epsilon^{\mu\nu\rho\sigma}
 n_{\mu} \bar n_{\nu}\epsilon^{T*}_{D,\rho}
 \epsilon^{T*}_{V,\sigma}A^p\bigg],
 \end{eqnarray}
where ${\cal A}^{s/p}=\xi_{int}^{s/p}+{\cal
M}_{int}^{s/p}+\xi_{exc}^{s/p}+{\cal M}_{exc}^{s/p}$.
$\xi_{int}^{s/p},\;{\cal M}_{int}^{s/p},\;\xi_{exc}^{s/p},\;{\cal
M}_{exc}^{s/p}$ correspond to the factorizable emission diagrams,
nonfactorizable diagrams, factorizable annihilation diagrams and
nonfactorizable diagrams respectively, and their analytic
expressions are given in Eq.(\ref{ints})-Eq.(\ref{mfp}).
$\epsilon^{T}_D$ and $\epsilon^{T}_V$ are the respective transverse
polarization vectors of $\bar D^*$ and vector mesons, $\bar n$ is
the light cone vector in which direction the momentum of the vector
meson is defined and $n$ is the opposite direction.
$\epsilon^{\mu\nu\rho\sigma}$ is the antisymmetric tensor with
$\epsilon^{0123}=1$.

The transversely polarized contributions
are suppressed by $r$ or $r_V$, and their expressions are given by
 \begin{eqnarray}
 \xi^s_{\rm int}(a_i)&=&8\pi C_Ff_Dr \int_0^1
 dx_1dx_3\int_0^{1/\Lambda}b_1db_1b_3db_3 \phi_B(x_1,b_1)
 \nonumber \\
 & &\times \{[-\phi_V^T(x_3)+r_V\left((x_3-1)\phi_V^a(x_3)+(x_3-3)\phi_V^v\right)]
 \nonumber\\
 & &\times a_i(t_i^{(1)})E_i(t_i^{(1)})h(x_1,(1-x_3)(1-r^2),b_1,b_3)S_t(x_3)
 \nonumber \\
 & &\ \ + r_V[\phi_V^a(x_3)-\phi_V^v(x_3)]a_i(t_i^{(2)}) E_i(t_i^{(2)})
 h(1-x_3,x_1(1-r^2),b_3,b_1)S_t(x_1)\}\;,
 \label{ints}
 \end{eqnarray}
 \begin{eqnarray}
 \xi^p_{\rm int}(a_i)&=&8\pi C_Ff_Dr \int_0^1
 dx_1dx_3\int_0^{1/\Lambda}b_1db_1b_3db_3 \phi_B(x_1,b_1)
 \nonumber \\
 & &\times \{[-\phi_V^T(x_3)-r_V\left((x_3-1)\phi_V^v(x_3)+(x_3-3)\phi_V^a\right)]
 \nonumber\\
 & &\times a_i(t_i^{(1)}) E_i(t_i^{(1)})h(x_1,(1-x_3)(1-r^2),b_1,b_3)S_t(x_3)
 \nonumber \\
 & &\ \ + r_V[\phi_V^a(x_3)-\phi_V^v(x_3)]a_i(t_i^{(2)}) E_i(t_i^{(2)})
 h(1-x_3,x_1(1-r^2),b_3,b_1)S_t(x_1)\}\;,
 \label{intp}
 \end{eqnarray}
 \begin{eqnarray}
  {\cal M}^s_{\rm int}(a_i)&=& 16\pi\sqrt{2N_c} C_F r \int_0^1
 [dx]\int_0^{1/\Lambda}b_1 db_1 b_2 db_2 \phi_B(x_1,b_1)\phi_D^T(x_2,b_2)
 \nonumber \\
 & &\times
 \left[-x_2\phi_V^T(x_3)a_i(t_d^{(1)})E_d(t_d^{(1)})h^{(1)}_d(x_i,b_i)\right.\nonumber\\
 &
 &\left.+\left(-r_V\big(\phi_V^a(x_3)+(2x_2+2x_3-5)\phi_V^v(x_3)\big)+(x_2-3)\phi_V^T(x_3)\right)
 a_i(t_d^{(2)})E_d(t_d^{(2)})h^{(2)}_d(x_i,b_i)\right]\;,
 \label{mds}
 \end{eqnarray}
 \begin{eqnarray}
 {\cal M}^p_{\rm int}(a_i)&=& 16\pi\sqrt{2N_c} C_F r \int_0^1
 [dx]\int_0^{1/\Lambda}b_1 db_1 b_2 db_2 \phi_B(x_1,b_1)\phi_D^T(x_2,b_2)
 \nonumber \\
 & &\times
 \left[-x_2\phi_V^T(x_3)a_i(t_d^{(1)})E_d(t_d^{(1)})h^{(1)}_d(x_i,b_i)\right.\nonumber\\
 &&\left.+\left(r_V\big(\phi_V^v(x_3)+(2x_2+2x_3-5)\phi_V^a(x_3)\big)+(x_2-3)\phi_V^T(x_3)\right)
 a_i(t_d^{(2)})E_d(t_d^{(2)})h^{(2)}_d(x_i,b_i)\right]\;,
 \label{mdp}
 \end{eqnarray}
 \begin{eqnarray}
 \xi^s_{\rm exc}(a_i)&=&8\pi C_Ff_B \int_0^1
 dx_2dx_3\int_0^{1/\Lambda}b_2db_2b_3db_3 \phi_{D}^T(x_2,b_2)
 \nonumber \\
 & &\times  \left[r r_V \bigg((x_2-1)\phi_V^a(x_3)+(x_2+1)\phi_V^v(x_3)\bigg)
 a_i(t_a^{(1)})E_a(t_a^{(1)}) h_a(1-x_3,x_2(1-r^2),b_3,b_2)S_t(x_3)\right.
 \nonumber \\
 & &\ \ \  \left.
 +\bigg(r
 r_V\big(x_3(\phi_V^v(x_3)-\phi_V^a(x_3))-2\phi_V^v(x_3)\big)+r^2\phi_V^T(x_3)\bigg)\right.\nonumber\\
 &&\left. \times a_i(t_a^{(2)})E_a(t_a^{(2)}) h_a(x_2,(1-x_3)(1-r^2),b_2,b_3)S_t(x_2)\right]\;,
 \label{excs}
 \end{eqnarray}
 \begin{eqnarray}
 \xi^p_{\rm exc}(a_i)&=&8\pi C_Ff_B \int_0^1
 dx_2dx_3\int_0^{1/\Lambda}b_2db_2b_3db_3 \phi_{D}^T(x_2,b_2)
 \nonumber \\
 & &\times  \left[-r r_V \bigg((x_2-1)\phi_V^v(x_3)+(x_2+1)\phi_V^a(x_3)\bigg)
 a_i(t_a^{(1)})E_a(t_a^{(1)}) h_a(1-x_3,x_2(1-r^2),b_3,b_2)S_t(x_3)\right.
 \nonumber \\
 & &\ \ \  \left.
 +\bigg(r r_V\big(x_3(\phi_V^v(x_3)-\phi_V^a(x_3))+2\phi_V^a(x_3)\big)+r^2\phi_V^T(x_3)\bigg)\right.\nonumber\\
 &&\left.\times a_i(t_a^{(2)})E_a(t_a^{(2)}) h_a(x_2,(1-x_3)(1-r^2),b_2,b_3)S_t(x_2)\right]\;,
 \label{excp}
 \end{eqnarray}
 \begin{eqnarray}
  {\cal M}^s_{\rm exc}(a_i)&=& 16 \pi\sqrt{2N_c} C_F  \int_0^1
 [dx]\int_0^{1/\Lambda}b_1 db_1 b_2 db_2
 \phi_B(x_1,b_1)\phi_{D}^T(x_2,b_2)
 \nonumber \\
 & &\times [2r r_V\phi_V^v(x_3)a_i(t_f^{(1)}) E_f(t_f^{(1)})h^{(1)}_f(x_i,b_i)\nonumber\\
 &&+(-x_2r^2\phi_V^T(x_3))a_i(t_f^{(2)})
 E_f(t_f^{(2)})h^{(2)}_f(x_i,b_i) ]\;.
 \label{mfs}
 \end{eqnarray}
 \begin{eqnarray}
  {\cal M}^p_{\rm exc}(a_i)&=& 16 \pi\sqrt{2N_c} C_F  \int_0^1
 [dx]\int_0^{1/\Lambda}b_1 db_1 b_2 db_2
 \phi_B(x_1,b_1)\phi_{D}^T(x_2,b_2)
 \nonumber \\
 & &\times [(-2 rr_V\phi_V^a(x_3)a_i(t_a^{(1)})
 E_f(t_f^{(1)})h^{(1)}_f(x_i,b_i)\nonumber\\
 &&+(-x_2r^2\phi_V^T(x_3))a_i(t_f^{(2)})
 E_f(t_f^{(2)})h^{(2)}_f(x_i,b_i) ]\;.
 \label{mfp}
\end{eqnarray}

 \section{Numerical results and discussions}
 \label{sec:numerical}

 \begin{table}[t]
\caption{Branching ratios of $B_{(s)}\to \bar{D}P$ decays calculated
in PQCD approach together with experimental data\cite{pdg} (unit:
$10^{-6}$). }
 \label{PPresults}
\begin{center}
\begin{tabular}{l c cc} \hline \hline
Modes &class & Experiments   &  Our results    \\
\hline
$\bar B^0 \to  D_s^- \pi^+$    &T   &$15.3\pm 3.5$      &$33.0_{-12.6-1.8-3.1}^{+17.4.+1.9+3.2}$   \\
$\bar B^0 \to  D_s^- K^+$        &E     &$$                 &$(0.47_{-0.15-0.12-0.04}^{+0.14+0.08+0.05})\times 10^{-2}$   \\
$\bar B^0 \to \bar D^0 \pi^0$      &C  &$$                 &$(4.50_{-1.6-1.2-0.41}^{+2.3+1.5+0.44})\times 10^{-2}$   \\
$\bar B^0 \to \bar D^0 \bar K^0$    &C &$$                 &$1.79_{-0.74-0.55-0.17}^{+0.95+0.78+0.18}$   \\
$\bar B^0 \to \bar D^0 \eta$        &C &$$                 &$(6.16_{-2.3-1.3-0.57}^{+3.1+1.5+0.61})\times 10^{-2}$   \\
$\bar B^0 \to \bar D^0 \eta^{\prime}$  &C  &$$             &$(4.12_{-1.5-0.90-0.38}^{+2.1+1.0+0.41})\times 10^{-2}$   \\
$\bar B^0 \to  D^- \pi^+$       &T      &$$                 &$1.10_{-0.43-0.07-0.10}^{+0.58+0.07+0.11}$   \\
$B^- \to \bar D^0 \pi^-$          &C   &$$                 &$0.17_{-0.06-0.03-0.02}^{+0.07+0.03+0.02}$   \\
$B^- \to \bar D^0 K^-$           &C    &$$                 &$2.89_{-0.95-0.81-0.27}^{+1.45+1.32+0.28}$   \\
$B^- \to D^- \pi^0$              &T     &$$                 &$0.75_{-0.27-0.06-0.07}^{+0.35+0.05+0.07}$   \\
$B^- \to D^- \bar K^0$            &A   &$<5.0$             &$0.19_{-0.06-0.16-0.02}^{+0.06+0.19+0.02}$   \\
$B^- \to D^- \eta$               &T     &$$                 &$0.37_{-0.14-0.03-0.03}^{+0.19+0.03+0.04}$   \\
$B^- \to D^- \eta^{\prime}$       &T    &$$                 &$0.24_{-0.10-0.02-0.02}^{+0.13+0.02+0.02}$   \\
$B^- \to D_s^- \pi^0$         &T        &$$                 &$17.9_{-6.8-1.01-1.7}^{+9.4+1.0+1.8}$   \\
$B^- \to D_s^- K^0$              &A    &$$                 &$(0.68_{-0.10-0.45-0.06}^{+0.39+0.90+0.07})\times 10^{-2}$   \\
$B^- \to D_s^- \eta$        &T          &$$                 &$10.3_{-3.7-1.4-0.96}^{+4.8+0.57+1.0}$   \\
$B^- \to D_s^- \eta^{\prime}$    &T     &$$                 &$5.71_{-2.28-0.46-0.53}^{+3.41+0.73+0.56}$   \\
\hline
$\bar B_s^0 \to  \bar D^0 \pi^0$    &E  &$$                 &$(9.54_{-3.4-1.7-0.89}^{+3.5+0.30+0.93})  \times 10^{-2}$ \\
$\bar B_s^0 \to  \bar D^0 K^0$      &C  &$$                 &$(2.43_{-1.0-0.98-0.23}^{+1.4+1.6+0.24})\times 10^{-2}$   \\
$\bar B_s^0 \to  \bar D^0 \eta$    &C   &$$                 &$1.10_{-0.41-0.26-0.10}^{+0.50+0.25+0.11}$   \\
$\bar B_s^0 \to  \bar D^0 \eta^{\prime}$   &C  &$$          &$2.58_{-0.94-0.54-0.24}^{+1.2+0.59+0.25}$   \\
$\bar B_s^0 \to  D^-  \pi^+$      &E    &$$                 &$0.19_{-0.06-0.03-0.01}^{+0.07+0.01+0.01}$   \\
$\bar B_s^0 \to  D^-  K^+$      &T       &$$                 &$1.21_{-0.43-0.13-0.11}^{+0.58+0.10+0.12}$   \\
$\bar B_s^0 \to  D_s^-  K^+$     &T      &$$                 &$29.4_{-10.6-3.5-2.7}^{+14.3+2.7+2.9}$   \\
 \hline \hline
\end{tabular}
\end{center}
\end{table}

 \begin{table}[t]
\caption{Branching ratios of $B_{(s)}\to \bar{D}V$ decays calculated
in PQCD approach together with experimental data\cite{pdg} (unit:
$10^{-6}$). }
 \label{PVresults}
\begin{center}
\begin{tabular}{l c cc} \hline \hline
Modes & Class  & Experiments   &  Our results    \\
\hline
$\bar B^0 \to  D_s^- \rho^+$      &T     &$<600$            &$35.9_{-13.2-1.9-3.3}^{+17.8+2.0+3.5}$   \\
$\bar B^0 \to  D_s^- K^{*+}$        &E   &$$                 &$(1.68_{-0.43-0.31-0.15}^{+0.63+0.42+0.16}) \times 10^{-2}$   \\
$\bar B^0 \to \bar D^0 \rho^0$      &C  &$$                 &$(3.44_{-1.3-0.78-0.32}^{+1.8+1.2+0.34}) \times 10^{-2}$   \\
$\bar B^0 \to \bar D^0 \bar K^{*0}$  &C &$<11.0$            &$1.92_{-0.77-0.52-0.17}^{+0.95+0.75+0.18}$   \\
$\bar B^0 \to \bar D^0 \omega$      &C  &$$                 &$(6.13_{-2.2-1.1-0.57}^{+2.7+1.6+0.59} )\times 10^{-2}$   \\
$\bar B^0 \to  D^- \rho^+$       &T      &$$                 &$1.27_{-0.48-0.06-0.12}^{+0.65+0.07+0.12}$   \\
$B^- \to \bar D^0 \rho^-$           &C  &$$                 &$(9.27_{-3.5-2.1-0.9}^{+4.6+3.5+0.91})\times 10^{-2}$   \\
$B^- \to \bar D^0 K^{*-}$           &C  &$$                 &$2.05_{-0.76-0.49-0.19}^{+1.1+0.94+0.20}$   \\
$B^- \to D^- \rho^0$             &T      &$$                 &$0.75_{-0.28-0.05-0.07}^{+0.37+0.05+0.07}$   \\
$B^- \to D^- \bar K^{*0}$       &A       &$$                 &$0.11_{-0.04-0.02-0.01}^{+0.03+0.01+0.01}$   \\
$B^- \to D^- \omega$               &T    &$$                 &$0.67_{-0.25-0.04-0.06}^{+0.32+0.03+0.06}$   \\
$B^- \to D_s^- \rho^0$             &T    &$$                 &$19.4_{-7.2-1.1-1.8}^{+9.5+1.0+1.9}$   \\
$B^- \to D_s^- K^{*0}$              &A  &$$                 &$(0.50_{-0.15-0.14-0.05}^{+0.16+0.14+0.05})\times 10^{-2}$   \\
$B^- \to D_s^- \omega$             &T    &$$                 &$16.8_{-6.2-0.93-1.6}^{+8.3+0.94+1.6}$   \\
$B^- \to D_s^- \phi$            &A       &$< 1.90$           &$0.13_{-0.04-0.02-0.02}^{+0.05+0.03+0.02}$   \\
\hline
$\bar B_s^0 \to  \bar D^0 \rho^0$   &E  &$$                 &$0.19_{-0.05-0.03-0.02}^{+0.05+0.04+0.02}$   \\
$\bar B_s^0 \to  \bar D^0 K^{*0}$    &C &$$                 &$(0.60_{-0.22-0.17-0.06}^{+0.32+0.27+0.06}) \times 10^{-2}$   \\
$\bar B_s^0 \to  \bar D^0 \omega$    &E &$$                 &$0.16_{-0.05-0.03-0.01}^{+0.05+0.02+0.01} $   \\
$\bar B_s^0 \to  \bar D^0 \phi$      &C &$$                 &$1.89_{-0.69-0.45-0.18}^{+1.0+0.71+0.18}$   \\
$\bar B_s^0 \to  D^-  \rho^+$       &E  &$$                 &$0.37_{-0.10-0.07-0.03}^{+0.11+0.06+0.04}$   \\
$\bar B_s^0 \to  D^-  K^{*+}$      &T    &$$                 &$1.42_{-0.48-0.12-0.13}^{+0.63+0.08+0.14} $   \\
$\bar B_s^0 \to  D_s^-  K^{*+}$     &T   &$$                 &$33.1_{-11.6-2.65-3.07}^{+15.4+1.63+3.06}$   \\
 \hline \hline
\end{tabular}
\end{center}
\end{table}

 \begin{table}[t]
\caption{Branching ratios of $B_{(s)}\to \bar{D^*}P$ decays
calculated in PQCD approach together with experimental data
\cite{pdg}(unit: $10^{-6}$). }
 \label{VPresults}
\begin{center}
\begin{tabular}{l c cc} \hline \hline
Modes&Class & Experiments   &  Our results    \\
\hline
$\bar B^0 \to  D_s^{*-} \pi^+$      &T          &$30.0\pm 7.0$      &$41.7_{-15.8-2.3-3.8}^{+21.9+2.5+4.1}$   \\
$\bar B^0 \to  D_s^{*-} K^+$           &E       &$$                 &$(0.36_{-0.09-0.10-0.03}^{+0.10+0.16+0.04})\times 10^{-2}$   \\
$\bar B^0 \to \bar D^{*0} \pi^0$        &C      &$$                 &$(4.19_{-1.61-1.16-0.39}^{+2.31+1.36+0.41})\times 10^{-2}$   \\
$\bar B^0 \to \bar D^{*0} \bar K^0$      &C     &$$                 &$2.35_{-0.93-0.61-0.22}^{+1.2+0.94+0.23}$   \\
$\bar B^0 \to \bar D^{*0} \eta$           &C    &$$                 &$(8.43_{-2.94-1.86-0.78}^{+3.86+2.04+0.82})\times 10^{-2}$   \\
$\bar B^0 \to \bar D^{*0} \eta^{\prime}$   &C   &$$                 &$(5.64_{-1.97-1.26-0.53}^{+2.59+1.37+0.55})\times 10^{-2}$   \\
$\bar B^0 \to  D^{*-} \pi^+$         &T         &$$                 &$1.21_{-0.49-0.07-0.12}^{+0.67+0.08+0.12}$   \\
$B^- \to \bar D^{*0} \pi^-$             &C      &$$                 &$(6.13_{-3.17-1.84-0.57}^{+4.29+2.33+0.60})\times 10^{-2}$   \\
$B^- \to \bar D^{*0} K^-$              &C       &$$                 &$0.71_{-0.36-0.39-0.07}^{+0.54+0.54+0.07}$   \\
$B^- \to D^{*-} \pi^0$               &T         &$<3.0$             &$0.65_{-0.25-0.05-0.06}^{+0.36+0.05+0.06}$   \\
$B^- \to D^{*-} \bar K^0$         &A            &$<9.0$             &$1.46_{-0.31-0.10-0.14}^{+0.44+0.04+0.14}$   \\
$B^- \to D^{*-} \eta$                &T         &$$                 &$0.53_{-0.19-0.05-0.05}^{+0.26+0.03+0.05}$   \\
$B^- \to D^{*-} \eta^{\prime}$       &T         &$$                 &$0.35_{-0.13-0.03-0.03}^{+0.17+0.02+0.03}$   \\
$B^- \to D_s^{*-} \pi^0$            &T          &$$                 &$22.7_{-8.7-1.2-2.2}^{+12.2+1.4+2.3}$   \\
$B^- \to D_s^{*-} K^0$            &A            &$$                 &$(6.95_{-1.65-1.41-0.65}^{+2.00+0.97+0.68})\times 10^{-2}$   \\
$B^- \to D_s^{*-} \eta$             &T          &$$                 &$8.94_{-3.56-0.93-0.83}^{+4.80+0.63+0.87}$   \\
$B^- \to D_s^{*-} \eta^{\prime}$     &T         &$$                 &$13.0_{-4.42-0.79-1.21}^{+5.87+0.78+1.27}$   \\
\hline
$\bar B_s^0 \to  \bar D^{*0} \pi^0$     &E        &          &$0.13_{-0.04-0.02-0.01}^{+0.05+0.02+0.01} $ \\
$\bar B_s^0 \to  \bar D^{*0} K^0$         &C      &          &$(3.25_{-1.3-1.1-0.3}^{+1.9+1.9+0.3})\times 10^{-2}$ \\
$\bar B_s^0 \to  \bar D^{*0} \eta$        &C     &          &$1.16_{-0.47-0.26-0.11}^{+0.57+0.21+0.11}$ \\
$\bar B_s^0 \to  \bar D^{*0} \eta^{\prime}$  &C   &          &$3.44_{-1.25-0.71-0.32}^{+1.53+0.71+0.33}$ \\
$\bar B_s^0 \to  D^{*-}  \pi^+$            &E     &          &$0.27_{-0.09-0.05-0.02}^{+0.09+0.04+0.03}$ \\
$\bar B_s^0 \to  D^{*-}  K^+$           &T        &          &$1.38_{-0.50-0.16-0.13}^{+0.66+0.11+0.13}$ \\
$\bar B_s^0 \to  D_s^{*-}  K^+$         &T        &          &$36.2_{-13.1-4.4-3.6}^{+18.7+3.4+3.8}$ \\
 \hline \hline
\end{tabular}
\end{center}
\end{table}

 \begin{table}[t]
\caption{Predicted branching ratios of $B_{(s)}\to \bar{D^*}V$
decays(unit: $10^{-6}$) and the percentage of transverse
polarizations ${\cal R}_T$ together with the experimental
data\cite{pdg}.}
 \label{VVresults}
\begin{center}
\begin{tabular}{l c c c c c} \hline \hline
Modes & Class & Experimental BRs  &  &  PQCD BRs  &${\cal R}_T$  \\
\hline
$\bar B^0 \to  D_s^{*-} \rho^+$     &T       &$$    &$$             &$68.2_{-25.2-5.24-6.34}^{+33.9+4.49+6.65}$  &$ (33_{-0.4-1.3}^{+0.4+1.4})\%$ \\
$\bar B^0 \to  D_s^{*-} K^{*+}$        &E     &$$    &$$             &$(1.91_{-0.63-0.50-0.18}^{+0.44+0.43+0.19})\times 10^{-2}$  &$ (40_{-0.0-6.9}^{+5.8+15.0})\%$ \\
$\bar B^0 \to \bar D^{*0} \rho^0$     &C     &$$    &$$             &$0.31_{-0.11-0.05-0.03}^{+0.13+0.04+0.03}$  &$ (84_{-0.3-3.4}^{+1.4+2.6})\%$ \\
$\bar B^0 \to \bar D^{*0} \bar K^{*0}$  &C   &$<40$  &$$             &$13.5_{-4.80-2.16-1.25}^{+6.20+1.76+1.32}$  &$ (81_{-0.6-4.5}^{+2.6+5.5})\%$ \\
$\bar B^0 \to \bar D^{*0} \omega$       &C   &$$    &$$             &$0.24_{-0.09-0.04-0.02}^{+0.11+0.27+0.02}$  &$ (72_{-1.4-8.5}^{+0.0+4.4})\%$ \\
$\bar B^0 \to  D^{*-} \rho^+$         &T      &$$    &$$             &$2.29_{-0.86-0.17-0.21}^{+1.13+0.18+0.22}$  &$ (34_{-0.7-1.6}^{+0.0+1.3})\%$ \\
$B^- \to \bar D^{*0} \rho^-$          &C     &$$    &$$             &$1.10_{-0.34-0.42-0.10}^{+0.42+0.15+0.11}$  &$ (88_{-2.1-3.3}^{+1.0+1.9})\%$ \\
$B^- \to \bar D^{*0} K^{*-}$        &C       &$$    &$$             &$26.1_{-7.90-4.88-2.43}^{+9.93+3.68+2.54}$  &$ (88_{-1.1-3.0}^{+1.4+2.8})\%$ \\
$B^- \to D^{*-} \rho^0$                &T     &$$    &$$             &$1.50_{-0.52-0.15-0.14}^{+0.70+0.11+0.15}$  &$ (42_{-1.5-1.5}^{+1.1+0.5})\%$ \\
$B^- \to D^{*-} \bar K^{*0}$            &A   &$$    &$$             &$2.25_{-0.55-0.31-0.21}^{+0.68+0.26+0.22}$  &$ (87_{-2.2-3.0}^{+0.0+0.8})\%$ \\
$B^- \to D^{*-} \omega$                &T     &$$    &$$             &$1.01_{-0.38-0.06-0.09}^{+0.52+0.07+0.10}$  &$ (25_{-1.5-4.0}^{+1.5+4.1})\%$ \\
$B^- \to D_s^{*-} \rho^0$              &T     &$$    &$$             &$36.8_{-13.6-2.69-3.42}^{+18.2+2.41+3.59}$  &$ (33_{-0.4-1.4}^{+0.4+1.5})\%$ \\
$B^- \to D_s^{*-} K^{*0}$                &A  &$$    &$$             &$0.11_{-0.02-0.03-0.01}^{+0.04+0.02+0.01}$  &$ (89_{-2.6-3.8}^{+1.4+0.18})\%$ \\
$B^- \to D_s^{*-} \omega$               &T    &$$    &$$             &$31.9_{-11.8-2.42-2.97}^{+15.7+2.09+3.11}$  &$ (33_{-0.6-1.6}^{+0.4+1.4})\%$ \\
$B^- \to D_s^{*-} \phi$                  &A    &$<12.0$    &$$        &$3.09_{-0.73-0.77-0.29}^{+0.86+0.55+0.30}$  &$ (89_{-2.0-2.9}^{+0.8+0.9})\%$ \\
\hline
$\bar B_s^0 \to  \bar D^{*0} \rho^0$    &E  &  &        &$0.17_{-0.04-0.03-0.02}^{+0.05+0.04+0.02}$   &$ (38_{-3.4-7.2}^{+4.3+12.6})\%$  \\
$\bar B_s^0 \to  \bar D^{*0} K^{*0}$  &C  &  &        &$0.49_{-0.16-0.08-0.05}^{+0.21+0.09+0.05}$   &$ (84_{-1.7-5.5}^{+0.99+3.9})\%$  \\
$\bar B_s^0 \to  \bar D^{*0} \omega$    &E  &  &        &$0.15_{-0.04-0.02-0.01}^{+0.04+0.02+0.02}$   &$ (37_{-3.1-6.8}^{+4.0+13.3})\%$  \\
$\bar B_s^0 \to  \bar D^{*0} \phi$    &C  &  &        &$11.7_{-4.08-2.37-1.09}^{+4.89+1.88+1.15}$   &$ (78_{-0.2-4.6}^{+1.5+5.1})\%$  \\
$\bar B_s^0 \to  D^{*-}  \rho^+$        &E  &  &        &$0.34_{-0.08-0.07-0.03}^{+0.10+0.06+0.03}$   &$ (39_{-2.4-8.1}^{+1.7+11.4})\%$  \\
$\bar B_s^0 \to  D^{*-}  K^{*+}$      &T   &  &        &$2.35_{-0.81-0.23-0.22}^{+1.06+0.18+0.23}$   &$ (32_{-0.57-1.6}^{+0.1+1.3})\%$  \\
$\bar B_s^0 \to  D_s^{*-}  K^{*+}$     &T  &  &        &$64.8_{-22.4-6.90-6.03}^{+29.6+5.57+6.32}$   &$ (32_{-0.2-1.5}^{+0.2-1.8})\%$  \\
 \hline \hline
\end{tabular}
\end{center}
\end{table}

Although the meson wave functions are not perturbatively calculable,
they are universal for all the decay channels. We can determine them
from the well measured decay channels, such as the $B\to D^{(*)} P$
decays \cite{BtoD1}. We use the same light cone distribution
amplitudes   as those we obtained in Ref.~\cite{BtoD1}. The
Lorentz structure of wave functions, decay constants,  and some
parameters of PQCD approach, are listed in the
appendix~\ref{wave-functions}. With the formulae we list in
Sec.~\ref{sec:formulae}, we can get the amplitude $\cal A$ for each
channel. The decay width for the $ B_{(s)} \to \bar D_{(s)}^*V$ mode
is given by
\begin{eqnarray}
 \Gamma &=&\frac{1}{16\pi m_B}(1-r^2)\sum_{\epsilon_D,\epsilon_V}|{\cal A}^X|^2  \;\;\;\;\;\;\;\;(X=N,T)\nonumber\\
        &=&\frac{1}{16\pi m_B}(1-r^2)(|{\cal A}^N|^2+2(|{\cal A}^s|^2+|{\cal A}^p|^2)).
 \label{decay width formula1}
\end{eqnarray}
For the other three kinds of decay modes $B\to \bar D_{(s)}^{(*)} P$
and $B\to \bar D_{(s)} V$, the decay width is given by
\begin{eqnarray}
\Gamma = \frac{1}{16\pi m_B}(1-r^2)|{\cal A}|^2.
 \label{decay width formula2}
\end{eqnarray}
With the decay width at hand, the branching ratio is given by
${\cal BR} = \Gamma \tau_B$. We take
 $\tau_{B^-}=(1.674\times10^{-12})s/\hbar$, $\tau_{\bar B^0}=(1.542\times10^{-12})s/\hbar$,
  $\tau_{\bar B^0_s}=(1.466\times10^{-12})s/\hbar$, and
  $G_F=1.16639\times10^{-5}\mbox{GeV}^{-2}$ \cite{pdg}. Our numerical results are listed in Table \ref{PPresults}, \ref{PVresults},
\ref{VPresults} and \ref{VVresults}. The first error is from the
hadronic parameters of $B_{(s)}$ meson wave functions(the decay
constants and the shape parameters): $f_B =(0.19 \pm
0.025)\mbox{GeV}$, $f_{B_s} =(0.24 \pm 0.03)\mbox{GeV}$,
$\omega_b=(0.40 \pm 0.05)\mbox{GeV}$ for $B$ meson, and
$\omega_b=(0.50 \pm 0.05)\mbox{GeV}$ for $B_s$ meson. The second
error arises from the choice of the hard scales, which vary from
$0.75t$ to $1.25t$, and the uncertainty of $\Lambda_{QCD}=(0.25\pm
0.05)\mbox{GeV}$. This uncertainty characterize the size of the
next-to-leading order QCD corrections, which is shown reasonable in
these tables.  The third error comes from the uncertainties of the
CKM matrix elements:
 \begin{eqnarray}
&&|V_{ub}| = 0.003\;59\pm 0.000\;16,\;\;
 |V_{cd}|= 0.225\;6\pm 0.001\;0,\;\;\nonumber\\
 &&|V_{cs}|= 0.973\;34\pm 0.000\;23,\;\; \gamma = {(77^{+30}_{-32})}^{\circ}.
\end{eqnarray}
It is easy to see that the most important theoretical uncertainty
comes from the non-perturbative hadronic parameters, which can be
improved later by the experiments.

At quark level, the decay channels related in this paper are all
$b\to u$  transitions. This type of decays are suppressed by the CKM
matrix elements, especially for the decays without a strange quark
in the final states. This is one reason why most of the decays have
small branching ratios with order $10^{-6}$ or $10^{-7}$. Another
reason is the absence of the color allowed emission diagrams with a
light meson emitted.

As stated in previous section, all these decays do not have
contributions from the penguin operators. For the tree operator
induced decays, we have only four type of topology diagrams
contributed: the color allowed diagrams (T), the color suppressed
diagrams (C), the W annihilation decays (A) and the W exchange
decays (E). All the decays are thus classified    in the tables.
From the numerical results, we can see that the pure annihilation
type (``W'' or ``E'') decay branching ratios are suppressed
comparing with the ``T'' or ``C'' emission diagrams dominant decay
channels. Remember that the ``T'' and ``C'' emission diagrams
dominant decay channels may also have ``W'' and ``E'' annihilation
type contributions, although they are suppressed relatively. Within
each category of decays, the  large differences between channels are
due to the Cabbibo suppression factor $V_{cd}/V_{cs}$.

   Usually   the nonfactorizable emission diagrams  are suppressed
   comparing the factorizable diagrams.
   The two nonfactorizable
diagrams in Fig.\ref{fig:int}(c,d) give nearly canceled
contributions if the  emitted meson is a light meson. However, it's
not the situation here when the $\bar D$ meson is emitted.  $\bar c$
quark and the light quark are very different in the emitted $\bar D$
meson. As a result, the nonfactorizable emission diagrams also give
non-negligible contributions. For example those channels with the
$\bar D^{(*)0}$ meson in the final state are   color suppressed. The
Wilson coefficients for factorizable contribution $\xi_{int}$ and
nonfactorizable  contribution ${\cal M}_{int}$ are $a_2=C_1+C_2/3$
and $C_2/3$, respectively. Since $a_2\approx C_2/3$, the
$\xi_{int}(a_2)$ and ${\cal M}_{int}(C_2/3)$ give similar
contributions.
 For these color suppressed modes, one can find that   the
annihilation diagrams can also give relatively large contributions.
  Our
numerical results indicate that sometimes the annihilation diagrams'
contributions do have the same order of magnitude as the emission
diagrams. We also find that the twist-3 distribution amplitudes play
an important role, especially in the annihilation diagrams.

 When the charged
$D^{(*)-}$ meson in the final state, the emission diagrams are
 the color favored ones, with the Wilson
coefficients $a_1=C_2+C_1/3$ and $C_1/3$, for the factorizable
diagrams and nonfactorizable diagrams,  respectively. In this
situation the nonfactorizable diagrams ${\cal M}_{int}(C_1/3)$ are
highly suppressed by the Wilson coefficient, comparing with the
factorizable diagrams $\xi_{int}(a_1) $, $\xi_{int}(a_1) >> {\cal
M}_{int}(C_1/3)$. This means that the dominant amplitudes with
$\xi_{int}(a_1)$ are nearly proportional to the product of $\bar
D^{(*)}$ meson decay constant and a B to light meson form factor.
 This type of $ B \to \bar DP$ decays have a little
smaller branching ratios than those corresponding $ B \to \bar DV$
decays, since the form factors of $B\to V$ are a little larger.
However, in the  color suppressed modes, $\xi_{int}(a_2) \sim {\cal
M}_{int}(C_2/3)$, or pure annihilation type decays, the above
conclusion is not satisfied. Because the large ${\cal
M}_{int}(C_2/3)$ or the annihilation diagrams will bring more
complicated situations. So the branching ratios of $B \to \bar{D}V$
are not definitely bigger than those of $B \to \bar{D}P$.  From the
  Table~\ref{PPresults} and \ref{PVresults},
 we can see that some branching ratios of $B \to \bar{D}P$ are bigger
 than those of $B \to \bar{D}V$, which is different from $b\to c \bar{u}q$
 decays.

For the $ B_{(s)} \to \bar D_{(s)}^*V$ decays, we also calculate the
ratio ${\cal R}_T$ of the transverse polarization in the branching
ratios, which is given by
\begin{eqnarray}
{\cal R}_T=\frac{2(|{\cal A}^s|^2+|{\cal A}^p|^2)}{|{\cal
A}^N|^2+2(|{\cal A}^s|^2+|{\cal A}^p|^2)}.
\end{eqnarray}
{From Eq.(\ref{ints})-Eq.(\ref{mfp}), we can find that the
transversely polarized contributions of the emission diagrams are
suppressed by the factor $r$, and those of the W exchange diagrams
are suppressed by the factor $rr_V$. That is the reason in category
``T'' and ``E'' decays, we have relatively small transverse
polarization factions.

For color suppressed emission diagrams (C),     the factorizable
contribution $\xi_{int}(a_2)$ and non-factorizable contribution
${\cal M}_{int}(C_2/3)$ are at the same order magnitude,    none of
which can give dominant contributions. The two diagrams of ${\cal
M}_{int}$ (longitudinal) cancel with each other and the two diagrams
of ${\cal M}_{int}^s$ and ${\cal M}_{int}^p$ (transverse) strengthen
with each other. So we can expect large transverse polarized
contribution of the branching ratio.} From table~\ref{VVresults}, we
can see that all the ``C'' type decays have transverse polarization
around 80\%.

For the W exchange type decays, the factorizable annihilation
contributions are suppressed by the Wilson coefficients, thus the
  dominant contribution is from nonfactorizable annihilation diagrams.
 One can find that the ratios of the transverse polarizations  for the W
exchange diagrams are around $40\%$. For   the W annihilation type
decays (A),  the factorizable diagrams dominate the branching ratios
due to the large Wilson coefficients. However, the two factorizable
annihilation diagrams strongly  cancel with each other in the
longitudinally polarized case, while   they strengthen with each
other in transversely polarized cases.
 In addition, there is also cancelations  between
the factorizable and nonfactorizable contributions for the
longitudinal polarizations. Therefore  the transverse polarizations
take a far more larger ratio in the branching ratios, which can be
as large as nearly $90\%$.

The transverse polarization ratio do not depend on the variation of
the CKM factors, since these kinds of overall factors cancel in the
ratio. The uncertainty shown in table~\ref{VVresults} are from the
hadronic uncertainty and factorization scale. Although these
uncertainties are small, it does not mean that the polarization
ratio is stable. In fact, it is quite sensitive to the hadronic wave
function shape of the final state meson \cite{vv} and the power
corrections.

$B^- \to D^0(\bar{D}^0) K^-$ decays can be used to measure $\gamma$
angle, see Ref.~\cite{Gronau:1998vg}, where the ratio
$r\equiv\frac{|A(B^- \to \bar{D}^0 K^-)|}{|A(B^- \to D^0 K^-)|}$ is
an important quantity. With the amplitudes we obtain, this ratio is
given as
 \begin{eqnarray}
r = \frac{|A(B^- \to \bar{D}^0 K^-)|}{|A(B^- \to D^0 K^-)|} =
0.092_{-0.003-0.003}^{+0.012+0.003}.
\end{eqnarray}
The first error comes from the choice of the hard scales, and the
second error comes from the CKM matrix elements. The uncertainty due
to the hadronic parameters are canceled in the ratio, thus the
calculation of the ratio is more precise and stable than the
individual channels. With the formulas in Ref.~\cite{Gronau:1998vg},
$\gamma$ can be measured experimentally through
\begin{eqnarray}
R_i\equiv\frac{2[\Gamma(B^+\to D_i K^+) +\Gamma(B^-\to D_i
K^-)]}{\Gamma(B^+\to \bar {D^0} K^+) +\Gamma(B^-\to D^0 K^-)}
=1+r^2\pm\sqrt{4r^2\cos^2\gamma-{\cal A}^2
\cot^2\gamma},\label{eq:R}
\end{eqnarray}
where $i=1,2$, $D_{1,2}=(D^0\pm\bar {D^0})/\sqrt{2}$  corresponds
to the two CP eigenstates, $R_i$ is defined as two charge-averaged
ratios for two CP eigenstates, and ${\cal A}={\cal A}_2-{\cal
A}_1$ with
\begin{eqnarray}
{\cal A}_i\equiv \frac{\Gamma(B^+\to D_iK^+)-\Gamma(B^-\to
D_iK^-)}{\Gamma(B^+\to \bar {D^0}K^+)+\Gamma(B^-\to D^0K^-)}.
\end{eqnarray}
In Ref.~\cite{Rdiagram}, authors give similar diagrams on the
relation of Eq.~(\ref{eq:R}). One can find that the sensitivity of
$\gamma$ to the other quantities increases as the $r$ decreases.
When $r\sim 0.1$ the extracted $\gamma$ will be very sensitive to
${\cal A}$. Thus our value of $r$ may be too small for the current
experiments.


 \section{Summary}
 \label{sec:summary}

In this paper, we investigate $B(B_s) \to \bar D_{(s)}P,\bar
D_{(s)}V,\bar D_{(s)}^*P,\bar D_{(s)}^*V$ decays under the framework
of perturbative QCD approach. We analyze the contributions of
different diagrams in the leading order approximation of the
$m_D/m_B$ expansion. It is found that the nonfactorizable emission
and annihilation diagram are also possible to give a large
contribution. However, the emission contributions are still dominant
in the branching ratios. All the branching ratios referred are
calculated and the ratios of the transversely polarized
contributions in the $B(B_s) \to \bar D_{(s)}^*V$ are also
estimated. We find that the transversely polarized contributions,
which mainly come from the nonfactorizable emission diagrams and
annihilation type diagrams, are very large. In some channels, they
are even dominant.  The branching ratios of this kind of decays are
around $10^{-6}$ and $10^{-7}$, which means the method of extraction
CKM angle $\gamma$ is not effective in experiments.

\section*{Acknowledgements}

This work is partly supported by National Natural Science
Foundation of China under Grants No. 10735080, No. 10625525, and
No. 10525523. We would like to thank W. Wang and Y.M. Wang for
fruitful discussions.

 \begin{appendix}

 \section{Wave functions and decay constants}
 \label{wave-functions}
\subsection{Wave functions of $B_{(s)}$ mesons}

The $B_{(s)}$ meson wave function are
  decomposed into the following Lorentz structures \cite{Bwave}
 \begin{eqnarray}
 &&\int\frac{d^4z}{(2\pi)^4}e^{ik_1\cdot z}\langle0|\bar
 b_{\alpha}(0)d_{\beta}(z)|B_{(s)}(P_1)\rangle\nonumber\\
 &=&\frac{i}{\sqrt{2N_c}}\left\{(\not P_1+M_{B_{(s)}})\gamma_5[\phi_{B_{(s)}}(k_1)-\frac{\not n-\not v}{\sqrt{2}}
 \bar\phi_{B_{(s)}}(k_1)]\right\}_{\beta\alpha}.
 \end{eqnarray}
  Here $\phi_{B_{(s)}}(k_1)$ and $\bar\phi_{B_{(s)}}(k_1)$ are the corresponding leading twist distribution
  amplitudes,
and $\bar\phi_{B_{(s)}}(k_1)$ contributes little, so we neglect
it. The final expression becomes
 \begin{eqnarray}
 \Phi_{B_{(s)}}=\frac{i}{\sqrt{2N_c}}{(\not{P_1}+M_{B_{(s)}})\gamma_5\phi_{B_{(s)}}(k_1)}.
 \end{eqnarray}
 The first determination of B-meson wave function was done in \cite{Bwave2}. In our work the distribution amplitude in the b-space is a little differently:
 \begin{eqnarray}
  \phi_{B_{(s)}}(x,b)&=&N_{B_{(s)}}x^2(1-x)^2\exp\left[-\frac{1}{2}(\frac{xM_{B_{(s)}}}
  {\omega_b})^2-\frac{\omega_b^2b^2}{2}\right],
  \end{eqnarray}
  Here b is the conjugate space coordinate of
  $\textbf{k}_{1\perp}$. $N_{B_{(s)}}$ is the normalization constant,
  which is determined by the normalization condition:
  \begin{eqnarray}
  \int^1_0 dx\phi_{B_{(s)}}(x,b=0)=\frac{f_{B_{(s)}}}{2\sqrt{2N_c}}.
  \end{eqnarray}

 \subsection{Wave functions and decay constants of light pseudoscalar mesons}

 The decay constant of the pseudoscalar meson is defined as:
 \begin{eqnarray}
 \langle 0|\bar
 q_1\gamma_{\mu}\gamma_5q_2|P(P_3)\rangle=if_PP_{3\mu}.
 \end{eqnarray}

 The Lorentz structure of light cone distribution amplitudes (for out-going state) for light pseudoscalar mesons is:
 \begin{eqnarray}
 &&\langle P(P_3)|q_{1\alpha}(0)\bar
 q_{2\beta}(z)|0\rangle\\ \nonumber
 &=&\frac{i}{\sqrt{2N_C}}\int_0^1dx
 e^{ixP_3\cdot z}\left[\gamma_5\not {P}\phi^A(x)+\gamma_5m_0\phi^P(x)+m_0\gamma_5(\not v\not
 n-1)\phi^T(x)\right]_{\alpha\beta}\;,
 \end{eqnarray}
 where  $v$ is the light cone direction along which the light pseudoscalar
 meson's momentum is defined, and $n$ is just opposite to it.
 The chiral scale parameter $m_0$ is defined as
 $m_0=\frac{M_P^2}{m_{q_1}+m_{q_2}}$.

 The distribution amplitudes are expanded by the
 Gegenbauer polynomials and their expressions are
 \begin{eqnarray}
 \phi_P^A(x)&=&\frac{3f_P}{\sqrt{2N_c}}x(1-x)\left[1+a_1^AC_1^{3/2}(t)+a_2^AC_2^{3/2}(t)
                                               +a_4^AC_4^{3/2}(t)\right],\\
 \phi_P^p(x)&=&\frac{f_P}{2\sqrt{2N_c}}\left[1+a_2^pC_2^{1/2}(t)+a_4^pC_4^{1/2}(t)\right],\\
 \phi_P^T(x)&=&-\frac{f_P}{2\sqrt{2N_c}}\left[C_1^{1/2}(t)+a_3^TC_3^{1/2}(t)\right],
 \end{eqnarray}
 with $t=2x-1$. The coefficients of the Gegenbauer polynomials are
 \cite{Pwave}
 \begin{eqnarray}
 a^A_{2\pi}&=& 0.44\;,\;a^A_{4\pi}=0.25\;,\;a^A_{1K}=0.17\;,\;a^A_{2K}=0.2\;,\nonumber\\
 a^p_{2\pi}&=& 0.43\;,\;a^p_{4\pi}=0.09\;,\;a^p_{2K}=0.24\;,\;a^p_{4K}=-0.11\;,\nonumber\\
 a^T_{3\pi}&=& 0.55\;,\;a^T_{3K}=0.35\;.
 \end{eqnarray}

 The decay constants are
\begin{eqnarray}
 f_{\pi} = 131 \mbox{MeV}
 \;,\; f_K = 160 \mbox{MeV}.
 \end{eqnarray}

 \subsection{Wave functions and decay constants of light vector mesons}

 The decay constants for the vector mesons are defined by
 \begin{equation}
 \langle 0|\bar q_1\gamma_\mu
 q_2|V(P_3,\epsilon)\rangle=f_Vm_V\epsilon_\mu,\;\;\; \langle 0|\bar
 q_1\sigma_{\mu\nu}q_2|V(P_3,\epsilon)\rangle =if^T_V(\epsilon_\mu
 P_{3\nu}-\epsilon_\nu P_{3\mu}).
 \end{equation}

 Up to twist-3 the distribution amplitudes are
 \begin{eqnarray}
 \langle V(P_3,\epsilon^*_L)|q_{1\alpha}(0)\bar
 q_{2\beta}(z)|0\rangle&=&-\frac{1}{\sqrt{2N_C}}
 \int_0^1dxe^{ixP_3\cdot z}\left[M_V\not\epsilon^*_L\phi_V(x)+\not\epsilon^*_L\not
 P_3\phi_V^t(x)+M_V\phi_V^s(x)\right]_{\alpha\beta},\nonumber\\
 \langle V(P_3,\epsilon^*_T)|q_{1\alpha}(0)\bar
 q_{2\beta}(z)|0\rangle&=&-\frac{1}{\sqrt{2N_C}}
 \int_0^1dxe^{ixP_3\cdot z}\left[M_V\not\epsilon^*_T\phi_V^v(x)+\not\epsilon^*_T\not
  P_3\phi_V^T(x)\right.\nonumber\\
  &&\left.+M_Vi\epsilon_{\mu\nu\rho\sigma}\gamma_5\gamma^\mu\epsilon^{*\nu}_Tn^\rho v^\sigma
  \phi_V^a(x)\right]_{\alpha\beta},
 \end{eqnarray}
 where $x$ is the momentum fraction of the $q_2$ quark. Contrary to the pseudoscalar
 case,
 here $n$ defines the light cone direction along which the momentum of light meson is taken
 and $v$ is just the opposite light cone direction.
 The twist-2 distribution amplitudes of vector mesons are defined as
 \begin{eqnarray}
 \phi_V(x)=\frac{3f_V}{\sqrt{2N_C}}x(1-x)\left[1+a^\parallel_1C_1^{3/2}(t)
                             +a^\parallel_2C_2^{3/2}(t)\right],\nonumber\\
 \phi_V^T(x)=\frac{3f_V}{\sqrt{2N_C}}x(1-x)\left[1+a^\perp_1C_1^{3/2}(t)
                             +a^\perp_2C_2^{3/2}(t)\right].\label{vwavef1}
 \end{eqnarray}
 and the corresponding values of the Gegenbauer moments are
 \cite{Vwave}
 \begin{eqnarray}
 a_{2\rho}^\parallel=a_{2\omega}^\parallel=0.15\pm0.07\;,\;a_{1K^*}^\parallel=0.03\pm0.02\;,\;
 a_{2K^*}^\parallel=0.11\pm0.09\;,\;a_{2\phi}^\parallel=0.18\pm0.08\;,\;\nonumber\\
 a_{2\rho}^\perp=a_{2\omega}^\perp=0.14\pm0.06\;,\;a_{1K^*}^\perp=0.04\pm0.03\;,\;
 a_{2K^*}^\perp=0.10\pm0.08\;,\;a_{2\phi}^\perp=0.14\pm0.07\;.\;
 \end{eqnarray}
 For the other distribution amplitudes, we use the asymptotic form
 \begin{eqnarray}
 &\phi_V^t(x) = \frac{3f_V^T}{2\sqrt{6}} t^2\;,\;&\phi_V^s(x)=\frac{3f_V^T}{2\sqrt{6}}(-t)\;,\nonumber\\
 &\phi_V^v(x) = \frac{3f_V}{8\sqrt{6}} (1+t^2)\;,\;&\phi_V^a(x)=\frac{3f_V}{4\sqrt{6}}(-t)\;.\;\label{vwavef2}
 \end{eqnarray}
The decay constants are listed below:
 \begin{eqnarray}
 f_{\rho} = 209 \pm 2 \mbox{MeV},\;\;
 f_{K^*}  = 217 \pm 5 \mbox{MeV},\;\;
 f_{\omega} = 195 \pm 3 \mbox{MeV},\;\;
 f_{\phi} = 231 \pm 4 \mbox{MeV},\;\;\nonumber\\
 f_{\rho}^T = 165\pm 9 \mbox{MeV},\;\;
 f_{K^*}^T = 185 \pm 10\mbox{MeV},\;\;
 f_{\omega}^T = 151\pm 9\mbox{MeV},\;\;
 f_{\phi}^T = 186 \pm 9 \mbox{MeV}.
 \end{eqnarray}

\subsection{Wave function of $D^{(*)}$ meson}

Up to twist-3 accuracy the two-particle light-cone distribution
amplitudes of $D^{(*)}$ meson are defined as
 \begin{eqnarray}
\langle D(P_2)| q_{\alpha}(z) \bar c_{\beta}(0)|0\rangle
&=&\frac{i}{\sqrt{2N_C}}
 \int_0^1dxe^{ixP_2\cdot z}\left[\gamma_5(\not P_2 + M)
 \phi_D(x,b)\right]_{\alpha\beta}\nonumber\\
 \langle D^{*}(P_2)|
 q_{\alpha}(z)\bar c_{\beta}(0)|0\rangle&=&-\frac{1}{\sqrt{2N_C}}
 \int_0^1dxe^{ixP_2\cdot z}\left[\not{\epsilon}_L(\not{P_2}+M_{D^*})\phi_{D^*}^L(x,b)
+\not{\epsilon}_T(\not{P_2}+M_{D^*})\phi_{D^*}^T(x,b)\right]_{\alpha\beta}\label{definition_of_D}
 \end{eqnarray}
 with
 \begin{eqnarray}
  \int_0^1 dx\phi_D(x,0)&=&\frac{f_D}{2\sqrt{2N_c}}\;,
  \int_0^1dx\phi_{D^*}^L(x,0)=\frac{f_{D^*}}{2\sqrt{2N_c}}\;,
  \int_0^1dx\phi_{D^*}^T(x,0)=\frac{f_{D^*}^T}{2\sqrt{2N_c}}\;,
  \end{eqnarray}
  as the normalization conditions.
  In the heavy quark limit we
  have
  \begin{equation}
  f_{D^*}^T-f_{D^*}\frac{m_c+m_d}{M_{D^*}}\sim
  f_{D^*}-f_{D^*}^T\frac{m_c+m_d}{M_{D^*}}\sim
  O(\bar\Lambda/M_{D^*}).
  \end{equation}
 Thus we will use $f_{D^*}^T=f_{D^*}$ in our calculation.
 The models for the distribution amplitude for D
 meson we used in this paper is
  \begin{eqnarray}
\phi_D^{(MGen)}(x,b) &=&
\frac{1}{2\sqrt{2N_c}}f_D6x(1-x)[1+C_D(1-2x)]\exp [\frac{-\omega^2
b^2}{2}].
 \end{eqnarray}
 The masses of $D_{(s)}^{(*)}$ meson we use are
 \begin{eqnarray}
m_{D}&=&1.869\mbox{GeV}, \qquad m_{D_s^-} = 1.968\mbox{GeV},  \nonumber \\
m_{D^{*}} &=& 2.010\mbox{GeV}, \qquad m_{D_s^{*-}} =
2.112\mbox{GeV}.
\end{eqnarray}
We take $f_D = 207 \mbox{MeV}$ \cite{Follana:2007uv}, $C_D = 0.5$,
$\omega = 0.1 \mbox{GeV}$ for $\bar{D}$ meson, $f_{D_s} = 241
\mbox{MeV}$\cite{Follana:2007uv}, $C_D = 0.4$, $\omega = 0.3
\mbox{GeV}$ for $\bar{D}_s$ meson. We use  the relations between
$f_D$ and $f_{D^*}$ derived from HQET \cite{Manohar:2000dt}:
\begin{eqnarray}
f_{D^{*}} = \sqrt{\frac{m_{D}}{m_{D^*}}}f_{D}, \qquad f_{D_s^{*-}}
= \sqrt{\frac{m_{D_s^-}}{m_{D_s^*-}}}f_{D_s^-}.
\end{eqnarray}

 \section{PQCD functions}
 \label{appendix:pQCDfunctions}

The pQCD functions appear in (\ref{int})-(\ref{mf}) and
(\ref{ints})-(\ref{mfp}) are listed as
 \begin{eqnarray}
 h_i(x_1,x_2,b_1,b_2)&=&K_{0}\left(\sqrt{x_1x_2}m_Bb_1\right)
 \nonumber \\
 & &\times \left[\theta(b_1-b_2)K_0\left(\sqrt{x_2}m_B
 b_1\right)I_0\left(\sqrt{x_2}m_Bb_2\right)\right.
 \nonumber \\
 & &\left.+\theta(b_2-b_1)K_0\left(\sqrt{x_2}m_Bb_2\right)
 I_0\left(\sqrt{x_2}m_Bb_1\right)\right]\;,
 \label{hd1}\\
 h_a(x_2,x_3,b_2,b_3)&=&\left(i\frac{\pi}{2}\right)^2
 H_0^{(1)}\left(\sqrt{x_2x_3}m_Bb_2\right)
 \nonumber \\
 & &\times\left[\theta(b_2-b_3)
 H_0^{(1)}\left(\sqrt{x_3}m_Bb_2\right)
 J_0\left(\sqrt{x_3}m_Bb_3\right)\right.
 \nonumber \\
 & &\left.+\theta(b_3-b_2)H_0^{(1)}\left(\sqrt{x_3}m_Bb_3\right)
 J_0\left(\sqrt{x_3}m_Bb_2\right)\right]\;,\label{hd2}
 \end{eqnarray}
 where $H^{(1)}(z)=J_0(z)+iY_0(z)$. And
 \begin{eqnarray}
 h^{(j)}_d&=&
 \left[\theta(b_1-b_2)K_0\left(Dm_B
  b_1\right)I_0\left(Dm_Bb_2\right) +\theta(b_2-b_1)K_0\left(Dm_B b_2\right)
 I_0\left(Dm_B b_1\right)\right]
 \nonumber \\
 &  & \times \left\{ \begin{array}{cc}
 K_{0}(D_{j}m_Bb_{2}) &  \mbox{for $D^2_{j} \geq 0$}  \\
 \frac{i\pi}{2} H_{0}^{(1)}(\sqrt{|D_{j}^2|}m_Bb_{2})  &
 \mbox{for $D^2_{j} \leq 0$}
 \end{array} \right\}\;,
 \label{hd}\\
 h^{(j)}_f&=& i\frac{\pi}{2}
 \left[\theta(b_1-b_2)H_0^{(1)}\left(Fm_B
 b_1\right)J_0\left(Fm_Bb_2\right) +\theta(b_2-b_1)H_0^{(1)}\left(Fm_B b_2\right)
 J_0\left(Fm_B b_1\right)\right]\;  \nonumber \\
 &  & \times \left\{ \begin{array}{cc}
 K_{0}(F_{j}m_Bb_{1}) &  \mbox{for $F^2_{j} \geq 0$}  \\
 \frac{i\pi}{2} H_{0}^{(1)}(\sqrt{|F_{j}^2|}m_Bb_{1})  &
 \mbox{for $F^2_{j} \leq 0$}
  \end{array} \right\}\;,\label{hf}
 \end{eqnarray}
 with the variables
 \begin{eqnarray}
 D^{2}&=&x_{1}(1-x_{3})(1-r^{2})\;,
 \nonumber \\
 D_{1}^{2}&=&(x_{1}-x_{2})(1-x_{3})(1-r^{2})\;,
 \nonumber \\
 D_{2}^{2}&=&(x_{1}+x_{2})r^{2}-(1-x_{1}-x_{2})(1-x_{3})(1-r^{2})\;,
 \nonumber \\
 F^{2}&=&x_{2}(1-x_{3})(1-r^{2})\;,
 \nonumber \\
 F_{1}^{2}&=&1-(1-x_2)(1-x_1-(1-x_{3})(1-r^2))\;,
 \nonumber\\
 F_{2}^{2}&=&x_2(x_1-(1-x_{3})(1-r^2))\;. \label{mis}
 \end{eqnarray}
The hard scales are determined by
 \begin{eqnarray}
 t_i^{(1)}&=&{\rm max}(\sqrt{(1-x_3)(1-r^2)}m_B,1/b_1,1/b_3)\;,
 \;\;\; t_i^{(2)}={\rm max}(\sqrt{x_1(1-r^2)}m_B,1/b_1,1/b_3)\;,
 \nonumber\\
 t_a^{(1)}&=&{\rm max}(\sqrt{x_2(1-r^2)}m_B,1/b_2,1/b_3)\;,\;\;\;
 t_a^{(2)}={\rm
 max}(\sqrt{(1-x_3)(1-r^2)}m_B,1/b_2,1/b_3)\;,\nonumber\\
 t_d^{(j)}&=&{\rm max}(Dm_B,\sqrt{|D_j^2|}m_B,1/b_1,1/b_2)\;,\;\;\;
 t_f^{(j)}={\rm
 max}(Fm_B,\sqrt{|F_j^2|}m_B,1/b_1,1/b_2)\;.\label{hardscale2}
 \end{eqnarray}
 Jet function appears in the factorization formulae is
 \begin{equation}
 S_t(x)=\frac{2^{1+2c}\;\Gamma(3/2+c)}{\sqrt{\pi}\;\Gamma(1+c)}[x(1-x)]^c,
 \end{equation}
with $c=0.5$ in this paper.

The expressions for the functions $E_l(t)$, with $l=i,a,d,f$, are
 \begin{eqnarray}
 E_i(t)&=&\alpha_s(t)\exp[-S_B(t)-S_3(t)]\;,
 \nonumber \\
 E_a(t)&=&\alpha_s(t)\exp[-S_{D}(t)-S_3(t)]\;,\nonumber\\
 E_d(t)&=&\alpha_s(t)\exp[-S(t)|_{b_3=b_1}]\;,
 \nonumber \\
 E_f(t)&=&\alpha_s(t)\exp[-S(t)|_{b_3=b_2}]\;,\label{Sudakov2}
 \end{eqnarray}
where the Sudakov exponent $S=S_B+S_{D}+S_3$, with $S_3$ as the
Sudakov exponent of the light meson. And the $S_j(x_i)$($j=B,D$ or
$3$) functions in Sudakov form factors are
 \begin{eqnarray}
 S_B(t)&=&s\left(x_1\frac{m_{B}}{\sqrt{2}},b_1\right)+2\int^t_{1/b_1}\frac{d\bar \mu}{\bar
 \mu}\gamma_q(\alpha_s(\bar \mu)),\\
 S_D(t)&=&s\left(x_2\frac{m_{B}}{\sqrt{2}},b_2\right)+2\int^t_{1/b_2}\frac{d\bar \mu}{\bar
 \mu}\gamma_q(\alpha_s(\bar \mu)),\\
 S_3(t)&=&s\left(x_3\frac{m_{B}}{\sqrt
2},b_3\right)+s\left((1-x_3)\frac{m_{B}}{\sqrt
2},b_3\right)+2\int^t_{1/b_3}\frac{d\bar \mu}{\bar
\mu}\gamma_q(\alpha_s(\bar \mu)),
 \end{eqnarray}
 with the quark anomalous dimension $\gamma_q=-\alpha_s/\pi$. The
explicit form for the  function $s(Q,b)$ is:
\begin{eqnarray}
s(Q,b)&=&~~\frac{A^{(1)}}{2\beta_{1}}\hat{q}\ln\left(\frac{\hat{q}}
{\hat{b}}\right)-
\frac{A^{(1)}}{2\beta_{1}}\left(\hat{q}-\hat{b}\right)+
\frac{A^{(2)}}{4\beta_{1}^{2}}\left(\frac{\hat{q}}{\hat{b}}-1\right)
-\left[\frac{A^{(2)}}{4\beta_{1}^{2}}-\frac{A^{(1)}}{4\beta_{1}}
\ln\left(\frac{e^{2\gamma_E-1}}{2}\right)\right]
\ln\left(\frac{\hat{q}}{\hat{b}}\right)
\nonumber \\
&&+\frac{A^{(1)}\beta_{2}}{4\beta_{1}^{3}}\hat{q}\left[
\frac{\ln(2\hat{q})+1}{\hat{q}}-\frac{\ln(2\hat{b})+1}{\hat{b}}\right]
+\frac{A^{(1)}\beta_{2}}{8\beta_{1}^{3}}\left[
\ln^{2}(2\hat{q})-\ln^{2}(2\hat{b})\right],
\end{eqnarray} where the variables are defined by
\begin{eqnarray}
\hat q\equiv \mbox{ln}[Q/(\sqrt 2\Lambda)],~~~ \hat b\equiv
\mbox{ln}[1/(b\Lambda)], \end{eqnarray} and the coefficients
$A^{(i)}$ and $\beta_i$ are \begin{eqnarray}
\beta_1=\frac{33-2n_f}{12},~~\beta_2=\frac{153-19n_f}{24},\nonumber\\
A^{(1)}=\frac{4}{3},~~A^{(2)}=\frac{67}{9}
-\frac{\pi^2}{3}-\frac{10}{27}n_f+\frac{8}{3}\beta_1\mbox{ln}(\frac{1}{2}e^{\gamma_E}),
\end{eqnarray}
$n_f$ is the number of the quark flavors and $\gamma_E$ is the Euler
constant. We will use the one-loop running coupling constant, i.e.
we pick up the four terms in the first line of the expression for
the function $s(Q,b)$.

 \end{appendix}


\end{document}